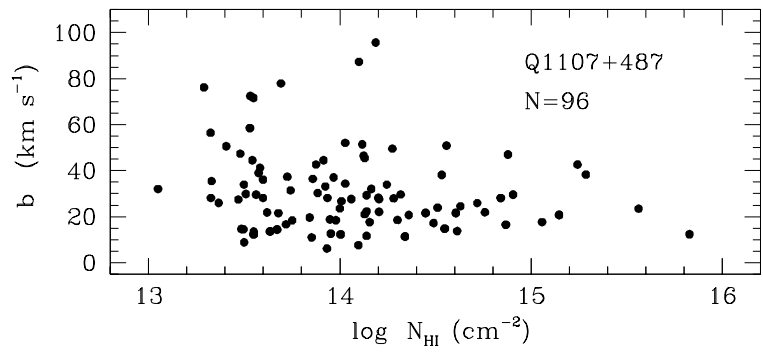

# The absorption spectra of Q1107+487 and Q1442+295

Ruth Carballo[1], Xavier Barcons[2] and John K. Webb[3]

[1] Departamento de Física Moderna, Universidad de Cantabria, Santander, Spain
[2] Instituto de Física, CSIC-Universidad de Cantabria, Santander, Spain
[3] School of Physics, University of New South Wales, Kensington, Australia

**Abstract.** We present the first moderate resolution ($\simeq$ 40–120 km s$^{-1}$) spectroscopic observations of the bright ($V \leq 16.7$) high-redshift QSOs Q1107+487 ($z_{em}$ = 2.965) and Q1442+295 ($z_{em}$ = 2.669) (Sanduleak & Pesch 1989). The relatively high signal to noise reached in the spectra along with an extensive wavelength coverage of the Ly$\alpha$ and the Ly$\beta$ forest allowed us to obtain, through profile fitting, column densities and Doppler parameters of the Lyman clouds towards these QSOs. The spectral coverage of regions longward of the Ly$\alpha$ emission line of the QSOs at the expected wavelengths of CIV $\lambda\lambda$1548,1550 at the redshifts of the Ly$\alpha$ forest allowed us to identify some heavy element absorption systems (hereafter HEASs) towards these QSOs. We have found no Lyman absorption system or HEAS towards these QSOs for which our data allow a deuterium measurement or to provide an interesting upper limit for the D/H ratio. The reason for this is that the Lyman lines with high column density detected towards these QSOs belong to absorption systems showing velocity structure.

## 1 Observations and data analysis

The spectra were obtained on five nights of 1991 January and March using the IPCS detector on the Intermediate Resolution Spectrograph at the 2.5 m INT at the Observatorio del Roque de los Muchachos (La Palma, Spain). For a detailed description of the data reduction, continuum fitting, generation of the line lists and techniques of profile-fitting of the absorption lines, the reader is refered to Carballo et al. (1995).

## 2 Lyman forest absorption systems

Only the Lyman forest systems with Ly$\alpha$ absorption longward of the QSO Ly$\beta$ emission were analised. When data at the corresponding Ly$\beta$ with reasonable S/N and spectral resolution were available and the lines were sufficiently unblended there, the Ly$\alpha$ and Ly$\beta$ lines were fitted simultaneously, in order to better constrain the $N_{HI}$, $b$ and $z$ parameters.

We have obtained the profile parameters of 96 Lyman forest lines towards Q1107+487 and 18 Lyman forest lines towards Q1442+295, within the redshift range $2.36 \leq z \leq 2.96$. The line sample towards Q1107+487 is large enough to obtain statistically significant information on the $b$ and $N_{HI}$ parameters of these clouds. The following results were obtained for the Ly$\alpha$ forest towards this QSO.

· At the relatively high S/N of our spectra no correlation between $b$ and $N_{HI}$ is apparent from the data (see Figure 1).

· The column density distribution of the Lyman clouds shows a flattening in the range $10^{13.6}\,\mathrm{cm}^{-2} \leq N_{HI} \leq 10^{14.1}\,\mathrm{cm}^{-2}$, above the estimated completeness limit of the sample ($10^{13.5}\,\mathrm{cm}^{-2}$), which we believe to be a real effect.

· The mean Doppler parameter of the Lyman lines is $\langle b \rangle = 31 \pm 2\,\mathrm{km\,s}^{-1}$ with a dispersion $\sigma(b) = 16 \pm 2\,\mathrm{km\,s}^{-1}$.

**Fig. 1.** Doppler parameter $b$ versus log $N_{HI}$ for the Lyman forest lines towards Q1107+487.

## 3    Heavy element absorption systems

A rather strong HEAS was found towards each QSO through the detection of the CIV doublet longward of the QSO Ly$\alpha$ emission line. The redshifts of the systems towards Q1107+487 and Q1442+295 are $z = 2.760$ and $z = 2.439$ respectively. Both systems show complex velocity structure, spanning total velocity intervals of $\simeq 270\,\mathrm{km\,s}^{-1}$, with a mean value of the velocity splittings between the different components of $\langle \Delta v \rangle = 161 \pm 23\,\mathrm{km\,s}^{-1}$. The system towards Q1442+295 has detected OI $\lambda$1302 and CII $\lambda$1334 absorption lines - apart from the CIV doublet - longward of the QSO Ly$\alpha$ emission line. Comparison of the OI, CII and CIV profiles for this system reveals a systematic change of the ionization state with velocity.